\documentclass[manuscript,nonacm]{acmart}
\AtBeginDocument{%
  }

\begin{document}

\title{A Personalized and Adaptable User Interface for a Speech and Cursor Brain-Computer Interface}


\author{Hamza Peracha}
\affiliation{%
  \institution{Department of Neurological Surgery, University of California, Davis}
  \city{Davis, CA}
  \country{USA}}
\email{hperacha@health.ucdavis.edu}

\author{Carrina Iacobacci}
\affiliation{%
  \institution{Department of Neurological Surgery, University of California, Davis}
  \city{Davis, CA}
  \country{USA}}
\email{ciacobacci@health.ucdavis.edu}

\author{Tyler Singer-Clark}
\affiliation{%
  \institution{Department of Neurological Surgery, University of California, Davis}
  \city{Davis, CA}
  \country{USA}}
\affiliation{%
  \institution{Biomedical Engineering Graduate Group, University of California, Davis}
  \city{Davis, CA}
  \country{USA}}
\email{tsingerclark@health.ucdavis.edu}

\author{Leigh R. Hochberg}
\affiliation{%
  \institution{School of Engineering and Carney Institute for Brain Sciences, Brown University}
  \city{Providence, RI}
  \country{USA}}
\affiliation{%
  \institution{VA RR\&D Center for Neurorestoration and Neurotechnology, VA Providence Healthcare}
  \city{Providence, RI}
  \country{USA}}
\affiliation{%
  \institution{Center for Neurotechnology and Neurorecovery, Department of Neurology, Massachusetts General Hospital, Harvard Medical School}
  \city{Boston, MA}
  \country{USA}}
\email{leigh_hochberg@brown.edu}

\author{Sergey D. Stavisky}
\affiliation{%
  \institution{Department of Neurological Surgery, University of California, Davis}
  \city{Davis, CA}
  \country{USA}}
\email{sstavisky@health.ucdavis.edu}

\author{David M. Brandman}
\affiliation{%
  \institution{Department of Neurological Surgery, University of California, Davis}
  \city{Davis, CA}
  \country{USA}}
\email{dmbrandman@health.ucdavis.edu}

\author{Nicholas S. Card}
\affiliation{%
  \institution{Department of Neurological Surgery, University of California, Davis}
  \city{Davis, CA}
  \country{USA}}
\email{nscard@health.ucdavis.edu}

\renewcommand{\shortauthors}{Peracha et al.}

\begin{abstract}
    Communication and computer interaction are important for autonomy in modern life. Unfortunately, these capabilities can be limited or inaccessible for the millions of people living with paralysis. While implantable brain-computer interfaces (BCIs) show promise for restoring these capabilities, little has been explored on designing BCI user interfaces (UIs) for sustained daily use. Here, we present a personalized UI for an intracortical BCI system that enables users with severe paralysis to communicate and interact with their computers independently. Through a 22-month longitudinal deployment with one participant, we used iterative co-design to develop a system for everyday at-home use and documented how it evolved to meet changing needs. Our findings highlight how personalization and adaptability enabled independence in daily life and provide design implications for developing future BCI assistive technologies.
\end{abstract}

\maketitle

\begin{figure*}[t]
  \centering
  \includegraphics[width=\textwidth]{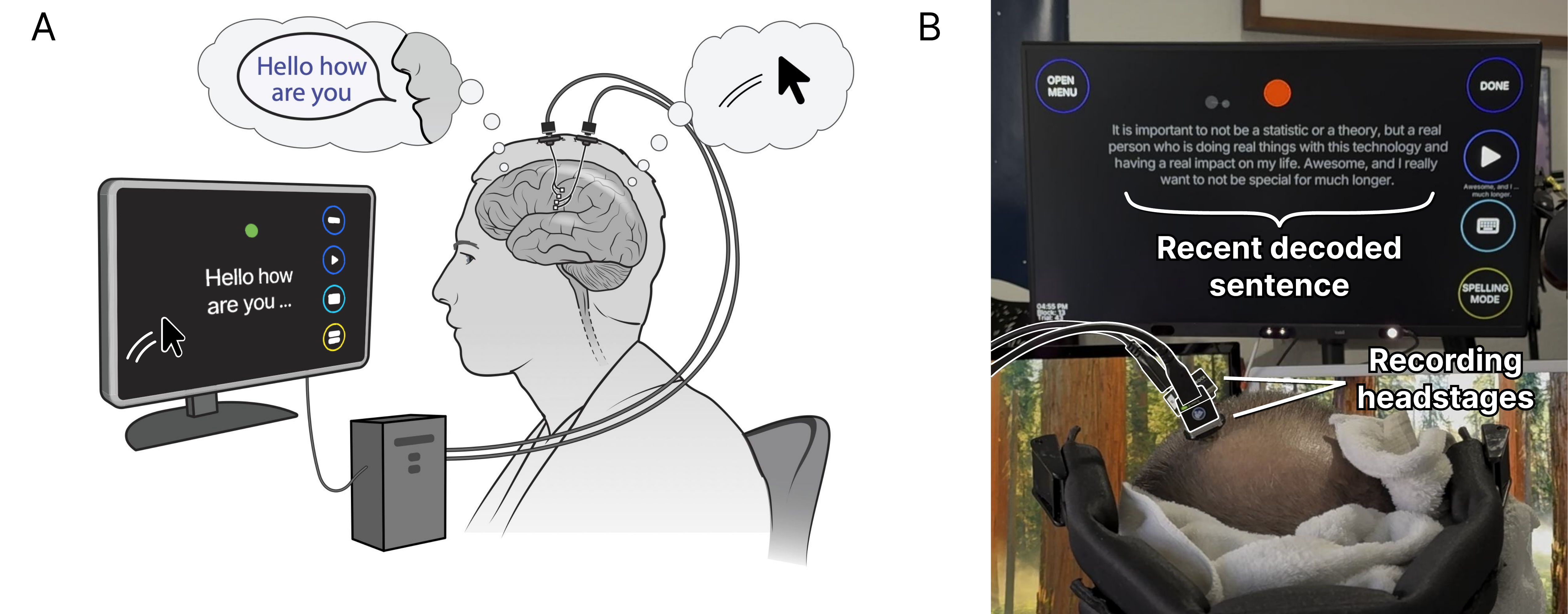}
  \caption{(A) Intracortical brain computer interface system schematic. Microelectrode arrays are surgically placed into the speech motor cortex. Neural information is transmitted through percutaneous wires to a computer system that decodes the neural data into words on a screen. (B) Participant T15 using the BCI system. T15’s personal computer monitors are pictured at the bottom, with the BCI systems monitor positioned on top, displaying the most recent decoded sentence.}
  \Description{A diagram of a person with implanted microelectrode arrays, along with an annotated image of participant T15 using the system. The diagram shows a person with four microelectrode arrays attempting to speak and move a cursor, and this is being reflected on the user interface of the computer in front of them. There is an image of this system in use, highlighting the neural recording headstages and the user interface displaying the attempted actions. Participant T15 is saying, "It is important to not be a statistic or a theory, but a real person who is doing real things with this technology and having a real impact on my life. Awesome, and I really want to not be special for much longer."}
\end{figure*}

\section{Introduction}
For people living with severe paralysis, the loss of voluntary motor control can have a profound impact on daily life by limiting movement, speech, and digital access. In the United States alone, about 5 million individuals live with paralysis \cite{armour2016prevalence}. For these individuals, maintaining the ability to interact with a computer can help them engage with work, entertainment, social connections, and other activities which promote independence and a higher quality of life \cite{antonino2019augmentative,londral2022assistive,fried2006purposes}. 

For many people living with paralysis, current solutions that enable communication and computer access are limited to alternative and augmentative communication (AAC) devices. AACs can range from low-tech options such as picture- and gaze-boards \cite{antonino2019augmentative,eyegazeboard}, to higher-tech options such as gyroscopic head mice or eye tracker based devices \cite{hamid202440,hou2024gazeswitch,etra2025}. However, existing AAC devices face challenges such as slow typing speeds, user fatigue, and difficulty in use due to the decline or loss of ocular motor function \cite{caligari2013eye,pannasch2008eye,edughele2022eye,proudfoot2016eye}.

Recent advances have demonstrated that brain-computer interfaces (BCIs) may be a viable AAC option. BCIs enable users to control devices by decoding intended movement from the brain, bypassing downstream injuries \cite{yuan2014brain,slutzky2019brain}. Implantable BCIs, such as those using intracortical microelectrode arrays, are surgically implanted and can directly record neural signals from within the brain, offering signals that are highly informative about a user’s movement intent. In previous work, implantable BCIs have been used to decode intended speech and cursor movements with high fidelity from individuals with severe paralysis, offering a new avenue for assistive technology and AACs \cite{card2024accurate,brandman2018robust,anumanchipalli2019speech,willett2023high,singer2025speech}. Previous work has also shown efforts to use this technology for computer interaction, including a cursor BCI for at home use \cite{weiss2019demonstration,simeral2021home}. Implantable BCIs have been studied for over two decades in a variety of first-in-human clinical trials (e.g., \cite{hochberg2006neuronal,collinger2013high,flesher2016intracortical,aflalo2015decoding,ajiboye2017restoration,anumanchipalli2019speech,pasley2012reconstructing,vansteensel2016fully}, see \cite{patrick2025state} for a more comprehensive list), and thanks to these proof-of-concept studies, there are now multiple neurotechnology companies seeking to create and validate medical-grade BCIs for people with paralysis \cite{swetlitz_2025,winkler_2025}.

While academic engineering labs have largely focused on developing new end-effectors and decoding algorithms for BCIs, relatively little has been explored about how users can best interact with these devices during daily use. This raises important questions on designing effective UIs that leverage this research to support BCI use for computer interaction and communication. In this work, we demonstrate a UI for a multimodal BCI system that enables real-time communication and full-featured control over a computer. BCI use is highly individualized as users vary in terms of abilities, preferences, and interaction methods. Our goal was to design a system that can be tailored to each user’s specific needs while remaining adaptable across a broader user base.

We propose the following two contributions:
\begin{enumerate}
    \item The design of a personalized BCI interface through a long-term co-design process with one user, demonstrating transferable design implications for an adaptable interface.
    \item A flexible system architecture with a shared backend and an adaptable user interface that allows for the system’s easy adaptation for different BCI users with diverse needs.
\end{enumerate}

\section{Related Work}
\subsection{Eye Gaze AAC Systems}
Eye gaze control is a popular interaction method for AAC and computer access \cite{hamid202440,hou2024gazeswitch,etra2025}, with commercial gaze systems such as the Tobii Dynavox (Sweden) becoming a common choice for users with paralysis \cite{TobiiDynavoxGlobal_2025}. Eye tracking technologies remain a viable choice for users with a wide range of neurological diseases, including late-stage amyotrophic lateral sclerosis (ALS), since ocular movements are often still intact despite limb and facial paralysis \cite{proudfoot2016eye}. These systems enable users to spell out words and control their personal computer using a cursor. Recent advances in gaze typing have significantly increased interaction speed by eliminating the need to dwell over each key \cite{kristensson2012potential}, addressing one of the limitations of dwell-based systems.

However, eye gaze systems still face significant limitations for computer use. Communication with these systems is slow \cite{pannasch2008eye,edughele2022eye}. Precise control over a computer cursor is also challenging, making navigation of interfaces with small buttons difficult, although there have been developments in design techniques to remedy this \cite{ashmore2005efficient,penkar2012designing}. Some users also face difficulties operating these systems due to irregularities in ocular movement, lack of reliability in outdoor environments, and requiring a direct line of sight \cite{caligari2013eye,proudfoot2016eye,evans2012collecting}.
\subsection{Switch-Based Systems}
Switch-based systems are another well-established access method. In switch-based systems, items are highlighted sequentially until the user activates a switch to select, allowing text entry and navigation. Although slower than direct selection methods and often more cognitively demanding \cite{simpson1999adaptive}, switch-based systems remain a robust interaction method that can be found in consumer devices and some BCI systems \cite{candrea2024click,appleswitch}.
\subsection{Non-Implantable BCI Systems}
Given the limitations of eye gaze devices, non-implantable BCIs using electroencephalography (EEG) have been explored as an alternative control method \cite{farwell1988talking,middendorf2000brain}. EEG-based BCIs remain relatively accessible and safe for anyone to use, and research has demonstrated their feasibility for basic computer control tasks and communication. P300-based spellers and steady-state visually evoked potential (SSVEP) systems have shown promise \cite{orhan2012rsvp,bin2009online}. Recent studies have demonstrated systems that are accurate and achieve communication rates upwards of 60 characters per minute in healthy subjects \cite{chen2015high,nakanishi2014high}, while research has also been focused on the usability of these systems \cite{guy2018brain,kubler2014user}.

Non-implantable BCIs face technical limitations. The resolution and signal-to-noise ratio of neural signals recorded by non-implantable BCIs are insufficient for high-fidelity control, limiting users to basic click or switch-based interfaces and selection tasks \cite{rashid2020current}. Although EEG-based BCIs remain promising as widely accessible AAC devices, they lack the signal-to-noise ratio required for the highest speed and accuracy in communication and computer use.
\subsection{Implantable BCI Systems}
Implantable BCIs have demonstrated a significant improvement in control and performance over non-implantable options. These systems can typically decode neural signals from within the skull with much higher resolutions and signal-to-noise ratios, enabling control modalities such as cursor control \cite{pandarinath2017high,weiss2019demonstration,bacher2015neural,dekleva2021generalizable,brandman2018rapid,singer2025speech}, robot arm control \cite{flesher2021brain,george2019biomimetic}, typing and handwriting interfaces \cite{jude2025intuitive,shah2024flexible,willett2021high}, and more recently, speech decoding \cite{card2024accurate,littlejohn2025streaming,metzger2023high,moses2021neuroprosthesis,wairagkar2025instantaneous,willett2023high,stavisky2025restoring,silva2024speech}. Previous work has described designing a user interface for at-home BCI systems but they lack a focus on the design process \cite{davis2022design,simeral2021home}.

Implantable BCIs face the obvious barrier of being less accessible due to the need for surgical implantation of the neural recording device, limiting their availability to participants in clinical trials until these devices are approved for broader use.
\subsection{User-Centered BCI Design}
BCIs benefit substantially from user-centered design. This requires a shift from focusing solely on metrics such as decoding performance towards the overall user experience through iterative design with end-users \cite{kubler2014user,schreuder2013user}. Recent work in communication-based BCIs has shown success in adopting participatory methods, using design-test-refine cycles while involving the end-users and domain experts \cite{gena2023bciai4sla}. However, these studies primarily focus on selection-based non-implantable BCI systems. To our knowledge, no prior work reports on long-term, iterative co-design with an implanted BCI system, and we build on these user-centered design practices to address our unique challenges.
\subsection{Speech Interfaces}
While prior work on speech BCIs has primarily focused on decoding accuracy, there is limited work on interface design. Commercial speech recognition software such as \textit{Dragon NaturallySpeaking} offers relevant precedents for a speech BCI \cite{dragon}, particularly regarding how users make corrections to errors in decoded sentences. Other work has explored error correction in speech recognition interfaces and how users can combine speech with other modalities to make corrections \cite{suhm2001multimodal}. These systems, however, require precise physical movements to use the software and make corrections, which are not possible by people with paralysis.

\section{Methods}
Design of new assistive technology presents unique challenges because user requirements cannot be easily determined through traditional methods \cite{wobbrock2011ability,newell2000user}. Additionally, Phillips and Zhao \cite{phillips1993predictors} emphasized that assistive technology may be abandoned if users' opinions are not considered during the design process. For emerging technologies such as implantable BCIs, these challenges are compounded as the technology’s capabilities and user control methods are still being explored.

We adopted an iterative co-design approach \cite{fraser2025shifting}, using our ongoing prototype as a technology probe \cite{hutchinson2003technology}. This allowed us to engage in participatory design, working collaboratively with the participant during the design process, while working versions of our prototype were deployed in the participant's home over an extended period of time. This approach allowed us to discover design requirements and feedback through continuous interaction that would not have been possible through traditional methods alone, such as larger-scale usability testing \cite{newell2000user}. We also applied Wobbrock et al.’s ability-based design principles \cite{wobbrock2011ability}, which focus on designing systems that adapt to users’ diverse abilities rather than requiring users to adapt to the system.
\subsection{Research Questions}
In this work, we address the following research questions:
\begin{enumerate}
    \item[\textbf{RQ1:}] How does a user with tetraplegia and severe dysarthria due to ALS integrate a speech and cursor BCI into daily life over an extended period of time?
    \item[\textbf{RQ2:}] What challenges arise during the implementation of the system, and how can iterative design address them?
    \item[\textbf{RQ3:}] What design implications emerge that can inform the design of future BCI systems?
\end{enumerate}
\subsection{Participant}
A 45-year-old man (T15) with tetraplegia and severe dysarthria due to ALS was enrolled in the BrainGate2 clinical trial (ClinicalTrials.gov number, NCT00912041) in 2023. T15 has four 64-microelectrode arrays placed in his left precentral gyrus, which is the part of the brain responsible for controlling the muscles involved in speech. Use of the system requires assistance from a member of the scientific team or a care partner to set up and remove the equipment.

Author D.M.B. is a board-certified neurosurgeon and is the site-responsible investigator for the BrainGate2 clinical trial at the University of California, Davis. He obtained informed consent, performed the operations, and continues to provide clinical care with participants while they are enrolled in the clinical trial.
\subsection{Study Design}
We evaluated the system through a 22-month-long longitudinal study with participant T15. Our approach combined continuous at-home deployment of new features along with periodic assessment of the BCI system, drawing from technology probe approaches that focus on real-world deployment and iterative design improvements \cite{hutchinson2003technology}.

We administered three surveys every six months: an assistive technology assessment evaluating the BCI’s overall effectiveness, a personal use task evaluation focused on an everyday computer task (we evaluated email composition), and a user-centered design questionnaire assessing the interface design and layout \cite{singer2021homeuse,gross2022design}. The user-centered design questionnaire was only administered at the second and third sessions. Initially, all interaction with T15 was done through a gyroscopic head mouse or via a skilled interpreter. Shortly after beginning this design process, all communication transitioned to being done through the BCI.
\subsection{Co-Design Process}
The co-design process involved continuous collaboration with T15 over 22 months through multiple channels. We conducted regular check-ins twice weekly, where T15 could provide brief feedback and discuss system usage. Additional focused design sessions were held periodically to review features and gather more detailed input. Between sessions, regular communication was maintained via text messages and shared documents, along with formal surveys run every 6 months.

Features were developed through ongoing collaboration with T15, including informal interactions, rapid prototype testing, and immediate deployment for real-world validation. T15 identified needs based on daily system usage, and the research team suggested new options when possible. Design decisions prioritized T15’s preferences and usage requirements.
\subsection{System Architecture}
Intracortical neural signals are transmitted to and decoded by a system of computers (Figure 1A) \cite{card2024accurate,singer2025speech,card2025long}. The computers run the BRAND software platform, which allows multiple Linux processes (“nodes”) across a distributed network optimized for real-time BCI control \cite{ali2024brand}. This distributed system enables rapid development with standardized frameworks for measuring system latency and jitter. 

The user interface consists of two nodes: one for logic and one for graphics display. The logic node is implemented as a finite state machine (FSM) (Figure 2), responsible for handling the different task states, such as idle, speaking, and so on (Figure 3), as well as receiving decoded neural signals in the form of text, cursor movements, or gestures. The graphics node uses Python 3.8 with the pyglet library (v2.0.12) \cite{pyglet}, which is made for designing visually-rich applications. Using this library, we created a node to allow users to interact with the BCI system. The graphics node consists of multiple "screens" (here we use the term “screen” to mean a visual configuration of user interface elements, as in “home screen”, and not to mean a physical display), with each screen corresponding to a different task state in the logic node. This system decoupling facilitates adding new features based on user feedback. The Python-based user interface is also ideal for development in a research environment, since most researchers are familiar with the language.

Users can control the interface using their preferred interaction method. Available input modalities include decoded speech, neural cursor control, decoded gestures, and eye gaze tracking. Decoded speech allows sentence formation for in-person communication or typing on a personal computer. Speech decoding uses a transformer-based decoder to convert neural signals to phoneme probabilities \cite{card2025long}. An n-gram language model then generates the most likely word sequences from these phonemes, which are fed to a large language model (OPT 6.7b) for additional rescoring in a fully local pipeline \cite{card2025long}. Neural cursor control enables BCI users to directly translate neural activity associated with attempted motor movements into precise cursor movements \cite{singer2025speech,hochberg2006neuronal,pandarinath2017high}. In our study, a time series of neural data is streamed into a cursor decoder and a click decoder \cite{singer2025speech, card2025long}. The cursor decoder is a linear model that maps neural signals to 2-D cursor velocities. The click decoder is a linear classifier that maps neural signals to probabilities of two discrete classes: "no action" vs "click". In our system, this enables pointer control on both the system's interface and the user's personal computer. Decoded gestures can be calibrated and mapped to cursor clicks or other discrete actions, such as scrolling. Eye gaze tracking provides an alternative means of interacting with the system’s interface. Together, these input modalities form the user-facing side of our BCI system, and users can recalibrate any input modality at any time via the menu.

For interfacing with the user's personal computer, we developed a desktop app that programmatically controls the keyboard and cursor. Users can paste decoded sentences into applications (Figure 3A) or enable neural cursor control to move and click the mouse on their personal computer.

\begin{figure}
  \centering
  \includegraphics[width=\linewidth]{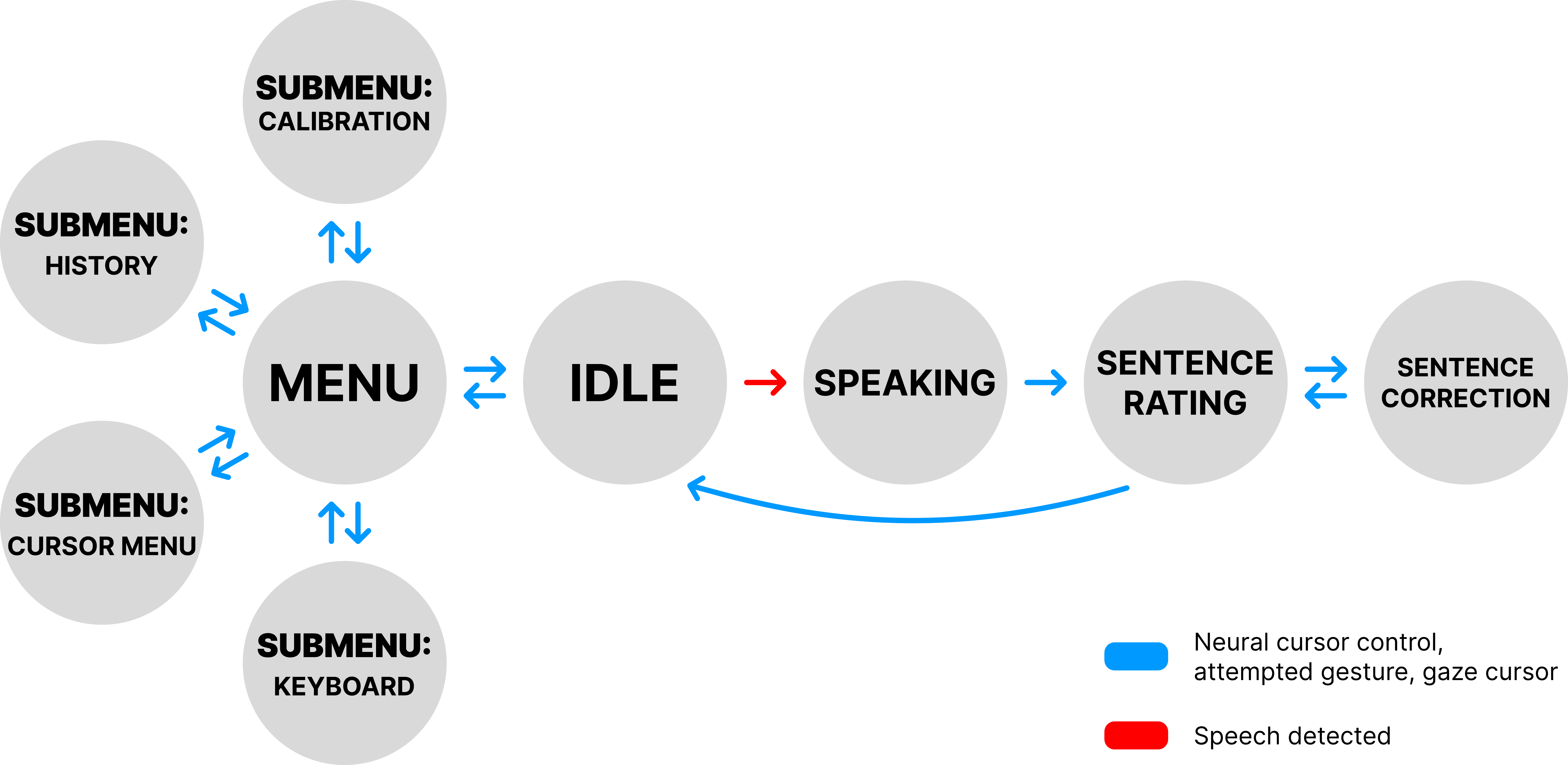}
  \caption{Finite state machine of the BCI interface for participant T15. It depicts different states and how the user navigates between them.}
  \Description{A diagram showing every task state in the finite state machine and how the user can navigate between them. Going from the ‘Idle’ task state to the ‘Speaking’ task state is triggered via detected attempted speech. Switching between all other task states can be done through neural cursor control, attempted gestures, or a gaze cursor.}
\end{figure}

\section{System Design}
This section outlines the design process of the BCI.
\subsection{T15 User Interface}
We report on the near-daily use of the BCI system by T15. T15 has used the system for over 4000 hours, for up to 19 hours per day, communicating up to 60 words per minute \cite{card2025long}. Using the system, T15 has maintained full-time employment, and is able to have conversations with his young child, despite being paralyzed \cite{bcivideo}. Over the course of 22 months, T15 has used the system through attempted speech, cursor control, gestures, and gaze tracking.

Given T15’s control methods, we were able to design a corresponding user interface as seen in Figure 3 and Figure 4. T15 typically uses the system sitting in his wheelchair, facing his personal desk in his bedroom. The interface is displayed on a 23.8” 1920x1080 resolution monitor mounted on an articulating arm, often positioned above T15’s personal computer monitors (Figure 1B).

The system begins in the “Idle” state (Figure 3A), displaying the most recently decoded sentence with quick actions such as play via text-to-speech, type to personal computer, and switch to spelling mode. When T15 begins speaking, the BCI automatically detects the attempted speech and transitions to the “Speaking” state (Figure 3B) \cite{card2024accurate}, displaying the decoded words in real time. When done speaking, T15 either selects the “Done” button or waits for a 6-second timeout.

The system then transitions to the “Sentence rating” state (Figure 3C), where T15 marks the correctness by selecting one of the four options: “Correct”, “One word wrong”, “Mostly correct”, and “Incorrect” which return T15 to the “Idle” state while communicating to the system and to others the correctness of the sentence. If corrections are needed, T15 can select the “Make corrections” button. Here, T15 is first presented with candidate sentences generated by the language model pipeline (Figure 3D). If the correct sentence is not among the candidates, T15 can select a specific word to view alternative word suggestions generated by ModernBERT (Figure 3E) \cite{modernbert}, add or delete words, refresh the suggestions, or manually type out corrections using an on-screen keyboard. Once corrections are complete, T15 returns to the “Sentence rating” state (Figure 3F) to mark the corrected sentence’s accuracy before returning to “Idle”. This gives the user the option to spend more time to ensure the sentence is fully correct if they choose to do so. In conversational exchanges, full sentence correctness is often not necessary to convey the user's intended message, which is where having the option to rate the sentence as mostly correct is useful. In contrast, a fully correct sentence may be desired when writing a report or typing a message.

The menu (Figure 3G) contains many additional features, including speech calibration (Figure 3I, Figure 3J), cursor calibration (Figure 4G, Figure 4H), eye tracker calibration, privacy mode (data is not saved while active), sentence history (Figure 3H displays the last 5 spoken sentences), additional personal computer controls (Figure 4F such as tab, enter, space, etc.), and language filtering (censors adult language). The on-screen keyboard can also be used to compose text directly without the use of the speech BCI, which is useful when constructing character strings that may not appear in a typical corpus, such as passwords and proper nouns. A list of select features and their deployment dates can be found in Table 1.

T15 can interact with the BCI interface and his personal computer via neural cursor or eye tracking. From the menu (Figure 4A), T15 can choose to control the BCI interface with neural cursor control (Figure 4B) or gaze control (Figure 4C). For personal computer control, T15 can choose to control the computer's cursor with neural cursor (4E) or gaze control (4F) through the menu (4D).

\begin{figure}
  \centering
  \includegraphics[width=\linewidth]{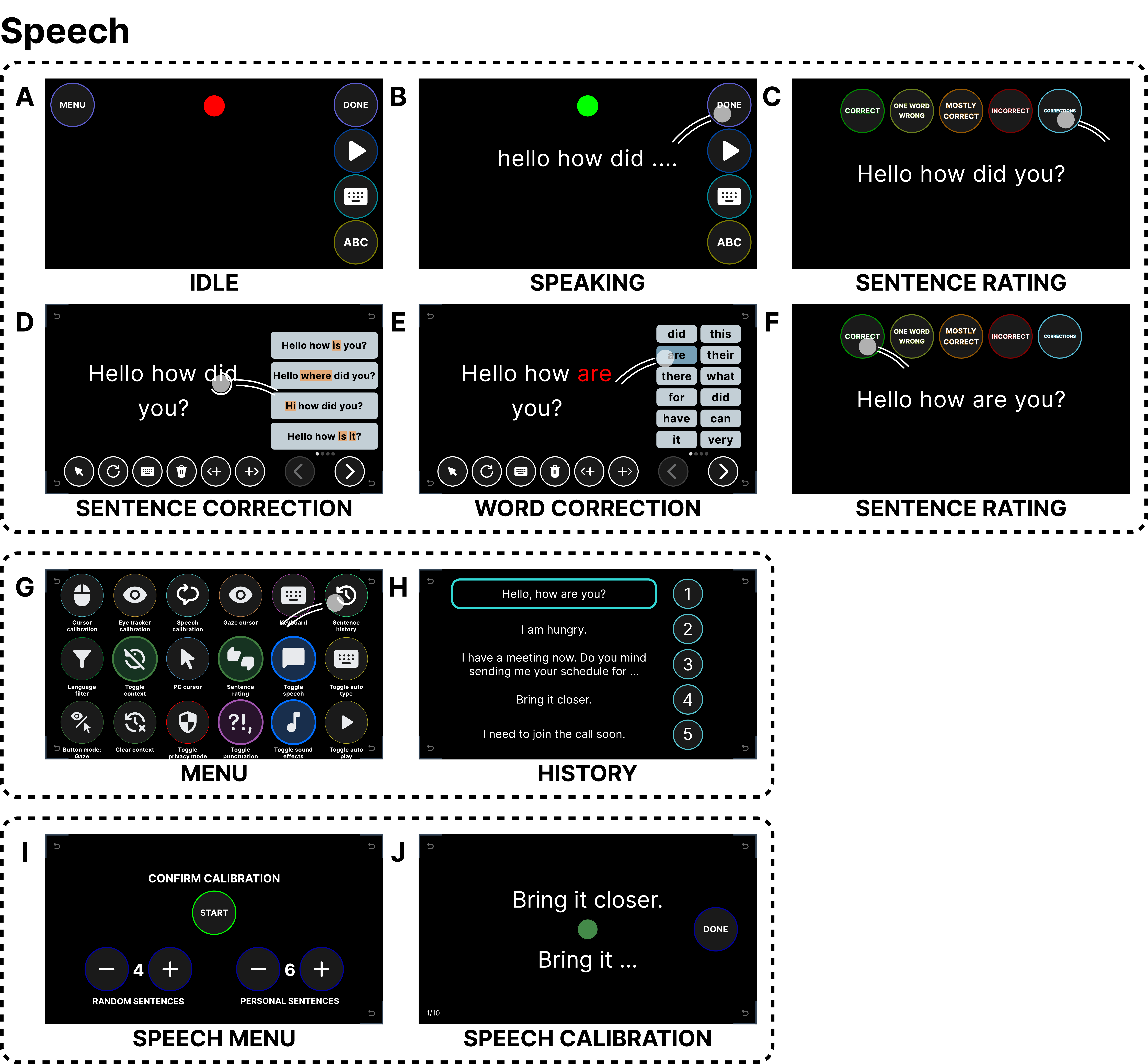}
  \caption{Speech-related user interface for T15. Subfigures (A)-(F) show a workflow of T15 speaking a sentence and making a correction. A gaze cursor pointer is shown interacting with the buttons on the screen. The following screens are shown: (A) idle (B) speaking (C) sentence rating (D) sentence correction (E) word correction with the selected word highlighted (F) sentence rating with the new corrected sentence. Subfigures (G) and (H) show how T15 can access the sentence history (H) through the menu (G). Subfigures (I) and (J) show accessing speech calibration (J) through the speech menu (I).}
  \Description{A breakdown of every speech-related screen of the user interface for participant T15. Subfigures (A) - (F) depict T15 speaking out "Hello how did you" and being corrected to "Hello how are you" using the word correction (F) screen. A history of sentences (H) is showing the last 5 sentences with "Hello, how are you?" currently selected. A speech calibration task is shown with a cue sentence at the top and the decoded sentence at the bottom (J).}
\end{figure}

\begin{figure}
  \centering
  \includegraphics[width=\linewidth]{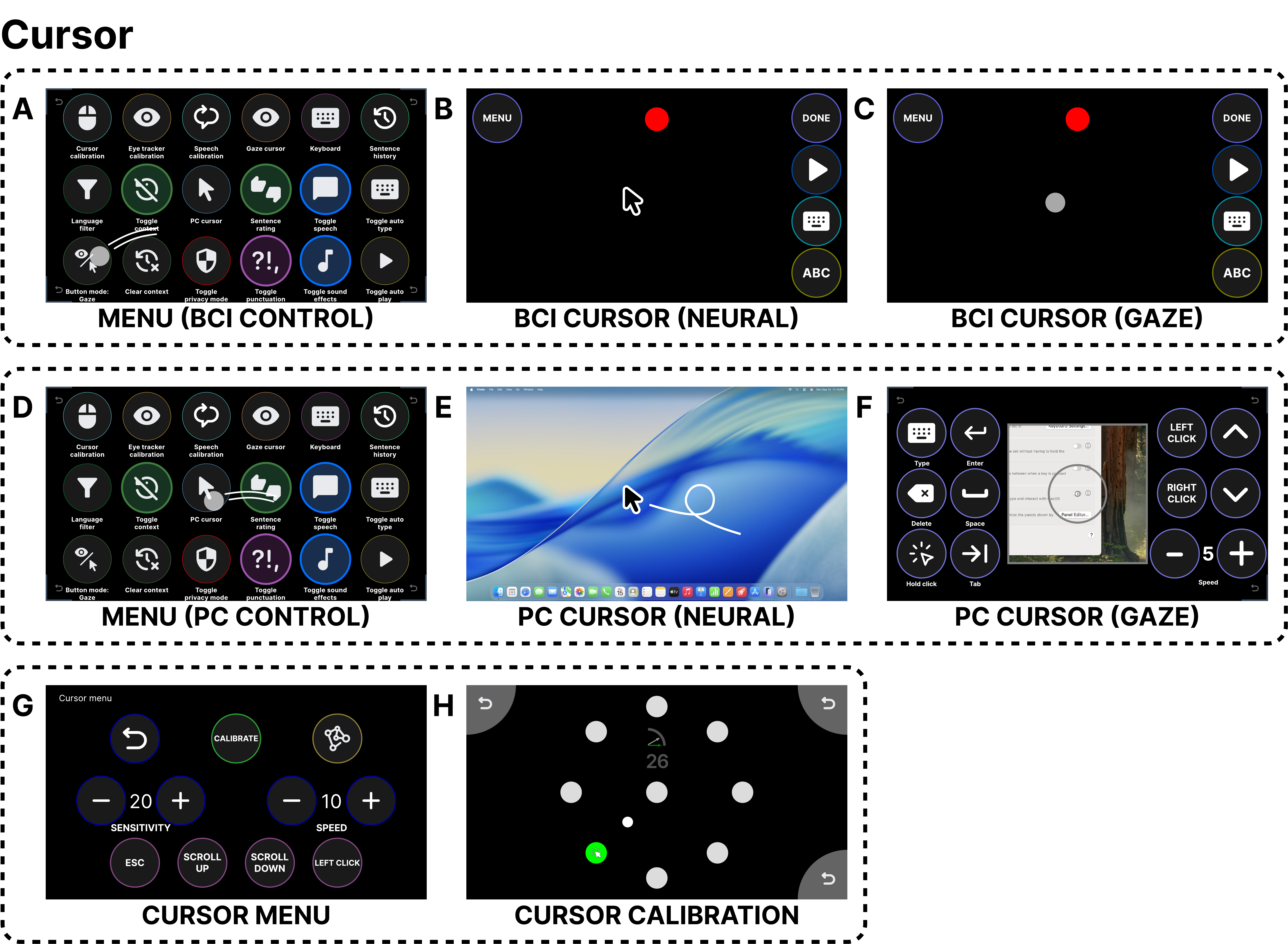}
  \caption{Cursor-related user interface for T15. Subfigures (A)-(C) show how T15 can switch between neural cursor control (B) and gaze control (C) for the BCI interface from the menu (A). Subfigures (D)-(F) show how T15 can use neural cursor (E) or gaze (F) to control his personal computer by selecting options from the menu (D). Subfigures (G) and (H) show accessing neural cursor calibration (H) through the cursor menu (G).}
  \Description{A breakdown of every cursor-related screen of the user interface for participant T15. Subfigures (A) - (C) show a cursor and a gaze pointer shown on the "idle" screen. (D) - (E) show cursor control on T15's personal computer with neural cursor and through a gaze based control panel. A cursor calibration game is shown in (H).}
\end{figure}

\begin{table*}[h]
  \caption{Features in T15's BCI user interface}
  \label{tab:features}
  \begin{tabular}{ccp{6cm}}
    \toprule
    Post-implant day&Feature name&Feature description\\
    \midrule
    147& Bluetooth keyboard& Allows the user to type decoded sentences on their personal computer.\\
    227& Sentence correction& Displays candidate sentences for a decoded sentence.\\
    232& Eye tracker button magnetization& Automatically guides the eye tracker cursor to the center of a button when the gaze enters a specific proximity, making dwell-based selection easier.\\
    358& Neural cursor control for personal computer& The user can control personal computer cursor via neural cursor control.\\
    364& Eye tracker calibration& The user can recalibrate the eye tracker through a menu option.\\
    462& Word-based correction& Candidate words are now shown for each word.\\
    462& Privacy mode& Does not save data while active.\\
    484& Miscellaneous updates& Language filter, sound effects, option to toggle features on and off, updates to UI.\\
    486& Neural cursor control in BCI& The user can control the BCI interface via neural cursor control.\\
    491& Refresh word suggestions& Refresh word suggestions and get new candidates.\\
    505& Word insertion and deletion& Insert and delete words in the word correction state.\\
    563& External cursor control& Allow a care partner to select buttons on the UI with a physical mouse.\\
    565& On-screen keyboard& On-screen keyboard that can be controlled with gaze or neural cursor.\\
    588& Aesthetic user interface changes& Redesign of the user interface.\\
    616& Updated word correction screen UI& Redesign of the sentence and word correction screens.\\
    654& Sentence history& Display the last 5 decoded sentences.\\
    668& Gaze-based cursor control panel& Control the personal computer cursor with gaze.\\
  \bottomrule
\end{tabular}
\end{table*}

\subsection{Adapting to Other Platforms}
By supporting multiple platforms, the system enables BCI use across a range of devices and in different environments, increasing flexibility and accessibility. Given the FSM-based design of our system, with one logic node and one graphics node, it is possible to swap out the graphics node to instead run the user interface on other devices. With this in mind, we designed versions of the BCI user interface for iPad (Figure 5A) and MacOS (Figure 5F). The same core functionality is displayed on these platforms, such as an idle screen, menu, sentence rating, and sentence correction. As long as the device is connected to the same local network that the BCI system is running on (via Ethernet, Bluetooth, or WiFi), it can read the task states and relevant information and display it accordingly. Cursor and keyboard control on the iPad is done through a Bluetooth device.

\begin{figure*}[t]
  \centering
  \includegraphics[width=\textwidth]{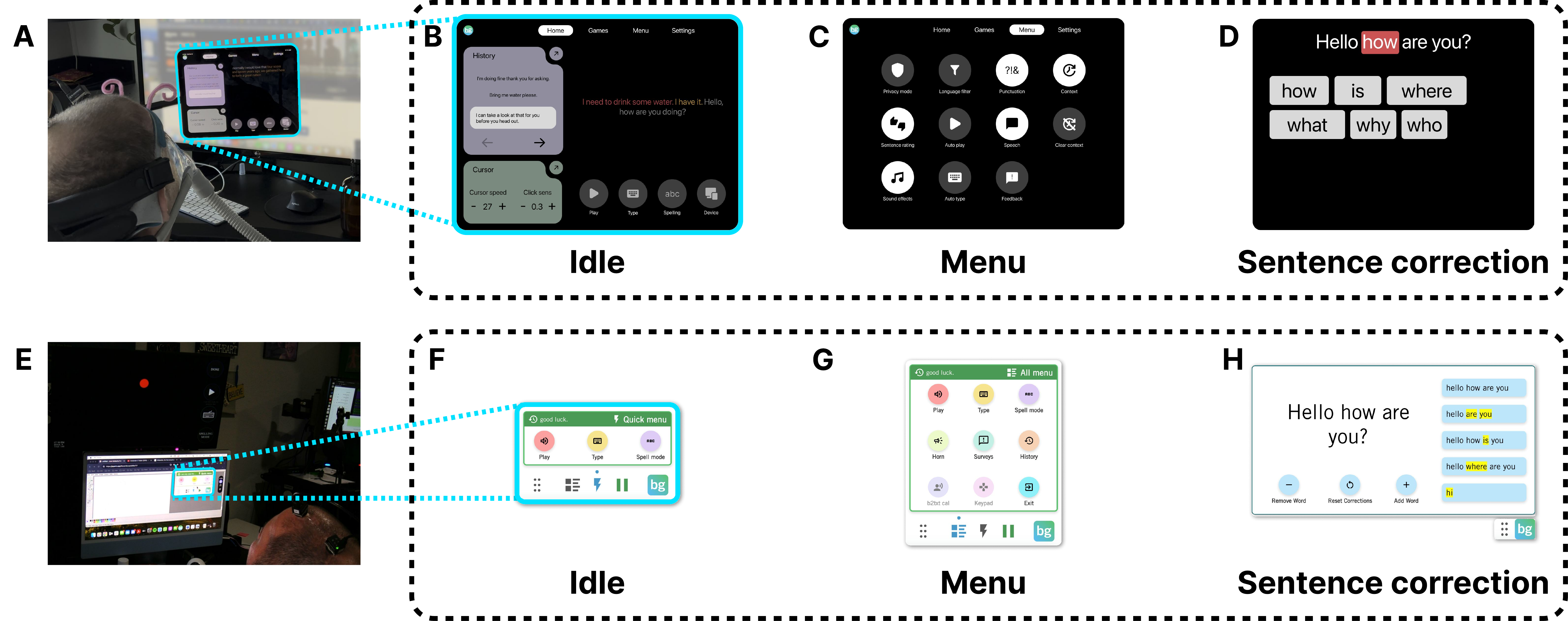}
  \caption{Participant T15 using the BCI on other platforms. (A) T15 using the BCI on an iPad. (E) T15 using the BCI on his iMac with a floating desktop application. Despite the differing form factor and operating systems, the applications follow the same structure as the original user interface. We show the following screens (B)(F) idle (C)(G) menu (D)(H) sentence correction.}
  \Description{A picture of participant T15 using the BCI system on his iPad. The interface is broken down into its key screens. The same is done for a floating desktop application running on his Mac.}
\end{figure*}

\section{Evaluation}
\subsection{Co-Design Insights}
A significant insight was the importance of sentence correction capabilities, as the speech decoder did not always produce a fully accurate output. Initial designs required users to respeak entire sentences if any words were decoded incorrectly, which proved to be impractical. Through continuous feedback, we developed three major iterations. The first presented full candidate sentence alternatives but lacked flexibility for performing single-word corrections. The second iteration provided candidate sentences and words on separate screens, which T15 found to be highly effective but revealed the need for a refined UI and supporting features. The final iteration presents sentence and word candidates on the same screen, with options to add or delete words, refresh suggestions, and manually type out words via an on-screen keyboard (Figure 3D, Figure 3E). While time spent in making corrections per trial increased from 19 to 34 to 62 seconds across iterations, the percentage of sentences marked as fully correct after making corrections increased substantially from 40\% and 41\% for the first two iterations to 59\% for the final iteration. This suggests that the iterative co-design process resulted in a corrections screen that T15 found worthwhile, prioritizing accuracy over speed when choosing to make corrections.

Additional adaptations from the co-design process included implementing dual control modalities (neural cursor and eye tracking), developing a layout optimized for eye tracking with large circular buttons and magnetization, integrating recalibration options for all input methods to ensure stable control over the system without external assistance, and adding a horn button to catch others' attention. T15's control method varied by context. Eye tracking may be preferable in appropriately-lit indoor environments with direct line of sight, while neural cursor control worked better in suboptimal lighting or viewing angles. This multimodal approach allowed T15 to leverage the strengths of eye tracking for interface navigation when conditions were favorable, while maintaining access to neural cursor control when needed, demonstrating that context-based control modality selection can address the limitations noted in prior work.

The multimodal interaction methods and flexible sentence correction options together are another key insight derived from this process. A user’s control over a BCI system will be highly dependent on their condition, the BCI hardware, and the decoding accuracy. This can vary not only from user to user but also for a single user due to factors such as day-to-day variability or disease progression. Control mode redundancies are important for enabling user independence and making the system more generalizable to other users with different needs \cite{wobbrock2011ability}.
\subsection{System Assessment}
T15 used the BCI consistently in his own home over 22 months, averaging 10 hours per day. The system supported a diverse number of daily activities, including communication with family, friends, and coworkers, email and text composition, phone and video calls, work-related tasks such as report writing, and leisure activities such as browsing the web and watching videos.

We administered three surveys at six-month intervals. Each survey consisted of Likert-scale questions with optional comment fields after each question. The personal use task evaluation focused on email composition, asking questions such as “Rate how frustrating it is to use email using your assistive technology system”. The assistive technology assessment survey evaluated the overall system as an assistive technology with questions such as “Rate how difficult it is to use your speech BCI system overall” and “Rate how independent you are when using your speech BCI system”. The user-centered design questionnaire (administered at sessions 2 and 3) assessed the interface design with questions including “Do you feel like your feedback impacts new system features?” and “Is the layout intuitive?”.

Survey results remained consistently positive throughout (Figure 6). T15 rated independence while using the system at a 5 and 4 (out of 5), with lower scores on independence during setup (Figure 6B), indicating high autonomy once started. Sustained satisfaction occurred despite the system evolving substantially between surveys, suggesting the co-design approach maintained user satisfaction while meeting changing needs. T15’s comments further supported independence as a key takeaway. T15 emphasized he could perform all computer tasks independently once the system is set up by a care partner, stating “[I] use it for hours myself” and “can do it all by myself”. This finding aligns with work from Phillips and Zhao \cite{phillips1993predictors}, showing independence as a key predictor of assistive technology adoption.

\begin{figure*}[t]
  \centering
  \includegraphics[width=\textwidth]{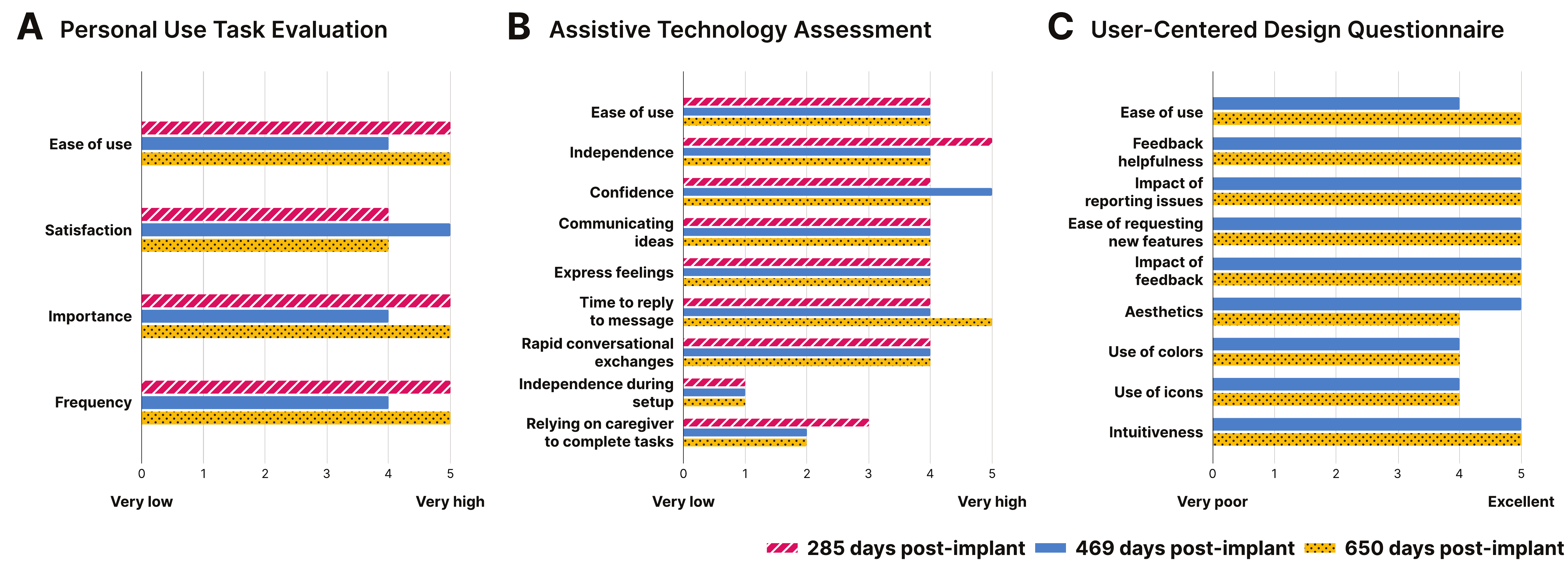}
  \caption{User interview data collected with participant T15 3 times over the course of one year.}
  \Description{Bar charts of each survey that was run with participant T15. Each survey's results remained positive on the Likert scale, except for requiring assistance to set up the BCI system. The scales for the personal use task evaluation and assistive technology assessment range from 1 (very low) to 5 (very high), and for the user-centered design questionnaire, from 1 (very poor) to 5 (excellent). Results for all questions averaged on a range of 4-5, except for reliance on a caregiver for setup (1) and for relying on a caregiver for completing tasks (2.3).}
\end{figure*}

\subsection{Longitudinal Usage Patterns}
System logs from T15’s 22-month use reveal insights into preferences and usage patterns over time.

T15 used two pointer control methods depending on the context. Eye tracking was preferred for the BCI interface, which was designed with large circular buttons and magnetization, making interaction more intuitive with gaze. Therefore, all interaction with the BCI system was done with eye tracking. The neural cursor offered more precision and was used for personal computer control, accounting for 11.5\% of the 4000+ hours of total system use time. The neural cursor was necessary for personal computer use as standard interface elements are not optimized for gaze and require finer control than eye tracking could provide.

Sentence and word correction played a key role in T15’s usage. T15 used word-level correction (Figure 3E) for 91.2\% of successfully corrected sentences, indicating its utility. A successfully corrected sentence is one that is marked as “correct” after entering the corrections screen. Sentence correction (Figure 3D) accounted for the other 8.8\%, and served as a faster option for shorter sentences, averaging 18 seconds in the corrections screen versus 46 seconds for word correction. Average sentence length was 9 words for sentence correction and 18 words for word correction. T15’s preference for word correction suggests that sentence-level alternatives, although faster, were often not sufficient, requiring more granular editing to achieve a fully correct sentence.

Within word correction, T15 had access to four editing features (Figure 3E): type word, delete word, add word before, and add word after. The delete function was used most often, while manually typing the word replacement out via an on-screen keyboard resulted in the highest correction accuracy (76\% of sentences were marked as fully correct when using this feature). Alternative editing options (adding words, deleting words, and refreshing candidate word suggestions) resulted in 65\% of sentences marked as fully correct. T15 may also choose not to fully correct a sentence to reduce the time spent making corrections.

\section{Discussion}
\subsection{Independence Through Personalization}
A key outcome of our longitudinal design process was that T15 remained satisfied with the system and was able to independently complete a wide variety of tasks, such as social interactions, leisure, and work, after initial system setup. We believe that these outcomes were a result of personalization of the system rather than just BCI decoding performance.

This process also enabled us to design for the changing needs of users. ALS is a progressive disease, and users' needs and control modalities may change over time. Therefore, the system should adapt accordingly. Examples of this flexibility in our system include T15 having the ability to control his BCI via neural cursor control or eye tracking. This adaptive capability is vital for maintaining long-term usability as the disease progresses and supports the user's abilities \cite{wobbrock2011ability}.

\subsection{Limitations and Future Considerations}
A clear limitation of our work is that the system has not been tested with a large number of users. Clinical trial inclusion criteria and the requirement of a surgically implanted BCI sensor result in a very limited user base. Due to this and the relatively recent emergence of intracranial neural speech decoding, large-scale testing is not yet feasible. Currently, this technology remains largely inaccessible outside of academic clinical trials, though recent efforts towards implanted BCI commercialization suggest that many more individuals may use such systems in the coming years.

Despite these limitations, this work provides value by addressing design questions that will become increasingly relevant as this technology becomes more widely available. The feedback collected can be carried forward to develop similar features for other users. Though the co-design process may not be feasible for large-scale studies or widespread deployment, here it proved to be a well-suited approach given the early stage of this field and the small, specialized user base. With this, we were able to apply 4 out of the 7 ability-based design principles outlined by Wobbrock at al. \cite{wobbrock2011ability}: ability, accountability, adaptation, and transparency (Table 2).

This approach has broader implications for developing adaptable user interfaces, especially in the field of assistive technology. Our findings suggest that multimodal redundancy supports independence despite variable BCI performance, correction options should be flexible depending on context of use, transparent control through calibration options and features such as sentence history maintain users' autonomy, and a shared backend architecture streamlines adaptation across users with different needs.

\begin{table}[h]
  \caption{Revisiting ability-based design principles.}
  \label{tab:principles}
  \begin{tabular}{ccp{8cm}}
    \toprule
    Principle&Applied&Details\\
    \midrule
    Ability& Yes& The system focuses on what users can do by providing multiple control modalities tailored to users' capabilities. This principle motivated integrating neural cursor control when eye tracking proved context-dependent.\\
    Accountability& Yes& When speech decoding is inaccurate, the system accommodates this through sentence corrections. Users can recalibrate at any time without external intervention. This principle drove the shift from external to user-initiated recalibration.\\
    Adaptation& Yes& The system adapts to users through multiple control modalities, alternative candidate sentences and words, and calibration options. This principle guided the three correction interface iterations, responding to limitations identified through T15's daily use.\\
    Transparency& Yes& Users maintain visibility and control over the system through features such as calibration, sentence history, and sentence rating and corrections.\\
    Performance& No& While the system logs usage data, it does not actively adapt to the user performance.\\
    Context& No& Users are able to manually change options such as control modalities, but the system does not proactively sense the context on its own.\\
    Commodity& No& This system requires a surgically implanted BCI sensor and specialized hardware and software to process the signals.\\
  \bottomrule
\end{tabular}
\end{table}

\section{Conclusion}
We report a highly personalized and adaptable user interface that enables users with paralysis to communicate and interact with a computer using an intracortical BCI, independently and in their own homes. We used an iterative co-design process with participant T15 that supported long-term daily use across communication, leisure, and work. This provides a step forward in BCI-based AAC solutions for communication and personal computer control that are both personalized and generalizable, offering a model for future user-centered BCI assistive technology development.

\begin{acks}
We would like to thank participant T15 and his care partners.

\smallskip

\noindent This work was supported by A.P. Giannini Foundation (https://ror.org/01e3cnp62), NIH-NIDCD (U01DC017844), VA RR\&D (A2295-R), ALS Association (24-AT-732), Neuralstorm NRT (2152260), ARCS Foundation, DP2 from the NIH Office of the Director and managed by NIDCD (1DP2DC021055), The Office of the Assistant Secretary of Defense for Health Affairs through the Amyotrophic Lateral Sclerosis Research Program (AL220043), Searle Scholars Program, and the Burroughs Wellcome Fund (https://ror.org/01d35cw23).
\end{acks}

\bibliographystyle{ACM-Reference-Format}
\bibliography{bibfile}

\end{document}